\documentclass[debug]{rmaa}

\usepackage{paralist}

\usepackage{psfrag,color}

\usepackage[latin1]{inputenc}

\def\p0{\phantom{0}}

\title{Radio continuum observations of LMC SNR J0550-6823} 

\author{
  L.~M.~Bozzetto,\altaffilmark{1} 
  M.~D.~Filipovi\'c,\altaffilmark{1}
  E.~J.~Crawford, \altaffilmark{1}
  J.~L.~Payne, \altaffilmark{1}
  A.~Y.~De~Horta \altaffilmark{1}
  and M.~Stupar \altaffilmark{2,3}}

\altaffiltext{1}{University of Western Sydney, Locked Bag 1797, Penrith South DC, NSW 1797, Australia}
\altaffiltext{2}{Department of Physics, Macquarie University, Sydney, NSW 2109, Australia}
\altaffiltext{3}{Australian Astronomical Observatory, PO Box 296, Epping, NSW 1710, Australia}

\shortauthor{Bozzetto et al.}
\shorttitle{LMC SNR J0550-6823}

\fulladdresses{
\item L.~M.~Bozzetto, M.~D.~Filipovi\'c, E.~J.~Crawford, J.~L.~Payne and A.~Y.~De~Horta: School of Computing and Mathematics, University of Western Sydney 2560, NSW, Australia
\item M.~Stupar: Department of Physics, Macquarie University, Sydney 2109, NSW, Australia
  }

\listofauthors{  L. M. Bozzetto,
  M. D. Filipovi\'c,
  E. Crawford, 
  J. L. Payne, 
  A. Y. De Horta,
  and M. Stupar}
\indexauthor{Bozzetto, L. M.}
\indexauthor{Filipovi\'c, M. D.}
\indexauthor{Crawford, E.}
\indexauthor{Payne, J. L.}
\indexauthor{De Horta, A. Y.}
\indexauthor{Stupar, M.}

\abstract{We report on new Australia Telescope Compact Array (ATCA) observations of the Large Magellanic Cloud (LMC) supernova remnant (SNR) J0550-6823 (DEM\,L328). This object is a typical horseshoe SNR with a diameter of $373\arcsec \times 282\arcsec \pm 4\arcsec$ ($90 \times 68 \pm 1$ pc), making it one of the largest known SNR{\bf s} in the Local Group. We estimate a relatively steep radio spectral index of \mbox{$\alpha = -0.79 \pm 0.27$}. However, its stronger than expected polarisation of {\bf $50\% \pm 10\%$} is atypical for older and more evolved SNRs. We also note a strong correlation between   [O{\sc iii}] and radio images, classifying this SNR as  oxygen dominant.}

\resumen{Este documento (\texttt{rm-journal-example.tex}---última
  actualización 9 de septiembre del 2007) proporciona un tutorial
  breve en el uso de la versión 3 de los macros de \LaTeX{}
  \texttt{rmaa} y además puede servir cómo modelo para la preparación
  de los artículos que se publicarán en la revista principal. Se puede
  encontrar más detalles en la guía del usuario
  (\texttt{authorguide.pdf}). Se supone que usted es ya familiar con
  los rudimentos del \LaTeX{}. En el caso contrario, se dan
  algunas referencias convenientes en el \texttt{authorguide.pdf}.}

\addkeyword{DEM\,L328}
\addkeyword{Large Magellanic Cloud}
\addkeyword{SNR J0550-6823}
\addkeyword{Supernova remnants}

\begin{document}
% Typeset article header
\maketitle

\section{Introduction}
\label{sec:intro}

Located at approximately 50~kpc \citep{2008MNRAS.390.1762D}, the Large Magellanic Cloud (LMC) is considered to be an ideal laboratory to study  Supernova Remnants (SNRs). Furthermore, the LMC is located in the direction of the South Celestial Pole, one of the coldest areas of the radio sky, making it possible to observe radio emissions without interference from galactic foreground radiation. Today's modern instruments make it possible to achieve detailed observations of these objects. 

There are over 50 well established SNRs in the LMC \citep{2010ApJ...725.2281K}  with an additional $\sim$20 SNR candidates (Bozzetto et al. in prep). This comprises one of the most complete samples of SNRs in external galaxies. Therefore, it is of prime interest to study LMC SNRs and compare them with their cousins in other galaxies such as M\,33 \citep{2010ApJS..187..495L}, M\,38 \citep{2010ApJ...710..964D}, the SMC \citep{2005MNRAS.364..217F,2007MNRAS.376.1793P,2008A&A...485...63F} and our Galaxy \citep{2008MNRAS.390.1037S, 2009BASI...37...45G}.

\citet{1976MmRAS..81...89D} observed an object in the LMC named DEM~L328 at H${\alpha}$ wavelengths and reported a nebulosity with a diameter of $25\arcmin \times 8\arcmin$. \citet{1976MNRAS.174..259S} detected this source in the Parkes 2700~MHz survey and reported a radio diameter of $\sim$3\arcmin. \citet{1998A&AS..127..119F}, using \emph{ROSAT} All Sky Survey (RASS) observations, detected X-ray emission from this source (LMC~RASS~309) and then calculated a spectral index from their Parkes data \citep{1998A&AS..130..421F} of $\alpha  = -0.37 \pm 0.06$.

Here, we present new medium-resolution observations of LMC~SNR~0550-6823. Observations, data reduction and imaging techniques are described in Section 2. The astrophysical interpretation of newly obtained moderate-resolution total intensity and polarimetric image are discussed in Section~3.

\section{Observations}
\label{sec:obs}

We observed SNR J0550-6823 with the Australia Telescope Compact Array (ATCA) on the  2$^\mathrm{nd}$ and 5$^\mathrm{th}$ of October 1997(project C634), using the array configuration EW375, at wavelengths of 3 and 6~cm ($\nu$=8640 and 4800~MHz). Baselines formed with the $6^\mathrm{th}$ ATCA antenna were removed from the imaging and the remaining five antennas were arranged in a compact configuration. Observations were taken in ``snap-shot'' mode, totalling $\sim$1.5 hours of integration over a 12 hour period. Source PKS~B1934-638 was used for primary calibration and source PKS~B0530-727 provided secondary (phase) calibration. The \textsc{miriad} \citep{MIRIAD} and \textsc{karma} \citep{Karma} software packages were used for reduction and analysis. It is well established that interferometers such as the ATCA will suffer from missing flux due to the missing short spacings. To compensate for this short-falling, we combined our new ATCA observations with  Parkes observations from \citet{1995A&AS..111..311F} and ATCA mosaic survey data from \citet{2005AJ....129..790D}. 

In addition to our own observations at 6~cm and 3~cm, we also used 73~cm ($\nu$=408~MHz) observations from \citet{1976AuJPA..40....1C}, 36~cm ($\nu$=843~MHz) observations from \citet{1984PASAu...5..537T} taken by Molonglo Observatory Synthesis Telescope (MOST) and 20~cm ($\nu$=1400~MHz) observations from the mosaics presented by \citet{2007MNRAS.382..543H} combining observations from ATCA and Parkes \citep{1995A&AS..111..311F}. We remeasured flux values for the 36~cm and 20~cm observations as shown in Table~1.

\begin{figure}[ht]
  \includegraphics[angle=-90, width=\columnwidth]{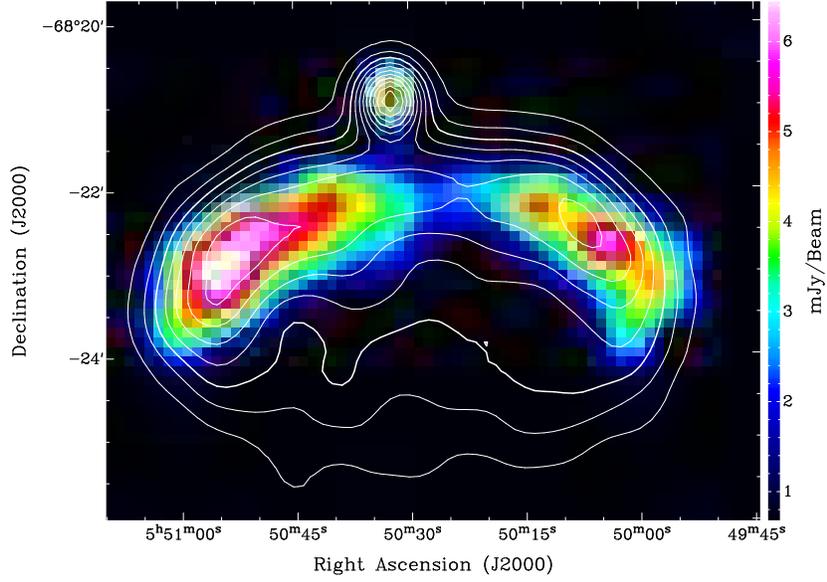}
  \caption{SNR J0550-6823 at 3~cm (8.6~GHz) overlaid with 6~cm (4.8~GHz) contours. The contours are from 1 to 21~mJy/beam in steps of 2~mJy/beam. The ellipse in the lower left corner represents the synthesised beam width (at 6~cm) of 40.2\,\arcsec$\times$35.3\arcsec. The sidebar quantifies the pixel map and its units are mJy/beam.}
  \label{3cm6cont}
\end{figure}

Our new images at 6 and 3~cm were initially created using only ATCA observations from project C634 and then processed using \textsc{miriad} multi-frequency synthesis \citep{1994A&AS..108..585S} and natural weighting. They were deconvolved using the {\sc clean} and {\sc restor} algorithms with primary beam correction applied using the {\sc linmos} task. A similar procedure was used for both \textit{U} and \textit{Q} Stokes parameter maps. Due to the low dynamic range\footnote {Defined as the ratio between the source flux and noise level.},  self-calibration could not be applied. The 6~cm image (Fig.~\ref{polar}) has a resolution of $36\arcsec \times 33\arcsec$  at PA=0$^\circ$ and an estimated r.m.s.~noise of 0.15~mJy/beam. This image was used for the polarisation study only. Similarly, we made an image of SNR~J0550-6823 at 3~cm, matching the resolution to the 6~cm image (Fig.~\ref{3cm6cont}). 

\begin{figure}[h]
  \includegraphics[angle=-90, width=\columnwidth]{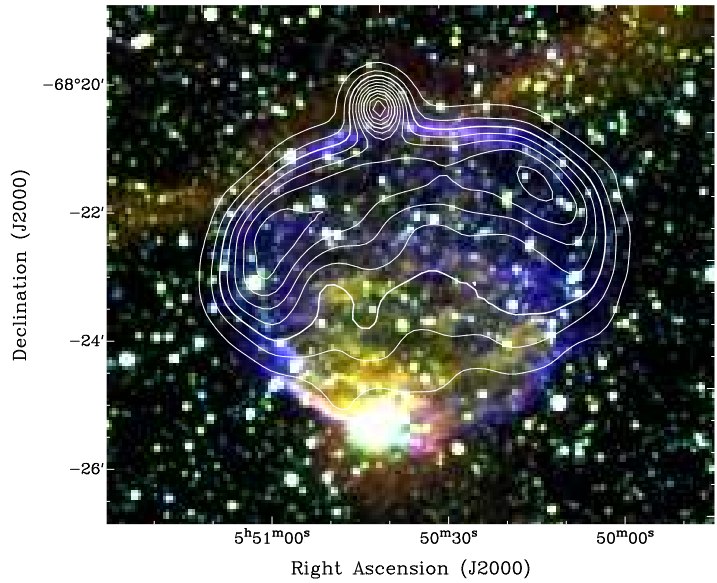}
  \caption{MCELS composite optical image \textrm{(RGB =H$\alpha$,[S\textsc{ii}],[O\textsc{iii}])} of SNR J0550-6823 overlaid with 6~cm contours from our new combined image of all ATCA and Parkes observations. The contours are from 1 to 21~mJy/beam in steps of 2~mJy/beam.}
  \label{mcels}
\end{figure}

Our analysis also made use of the Magellanic Cloud Emission Line Survey (MCELS) by \citet{mcels}.  This survey was  carried out with the 0.6-m University of Michigan/CTIO Curtis Schmidt telescope, equipped with a SITE $2048 \times 2048$\ CCD having a field of 1.35\arcdeg\ at a scale of 2.4\arcsec\,pixel$^{-1}$. They mapped both the LMC and SMC  in narrow bands corresponding to H${\alpha}$, [O\,{\sc iii}] ($\lambda$=5007\,\AA) and [S\,{\sc ii}] ($\lambda$=6716,\,6731\,\AA), matching red and green continuum bands in order to subtract most of the stars from the images to reveal the full extent of the faint diffuse emission. All of the data have been flux-calibrated and assembled into mosaic images, a small section of which is shown in Fig.~\ref{mcels}. Further details regarding the MCELS are given by Smith et al. (2006) and at http://www.ctio.noao.edu/mcels. Here, for the first time, we present optical images of this object in combination with our new radio-continuum data.

\section{Results and Discussion}

\begin{figure}[h]
 \includegraphics[angle=-90, width=110pt, trim=90pt 0pt 0pt 0pt]{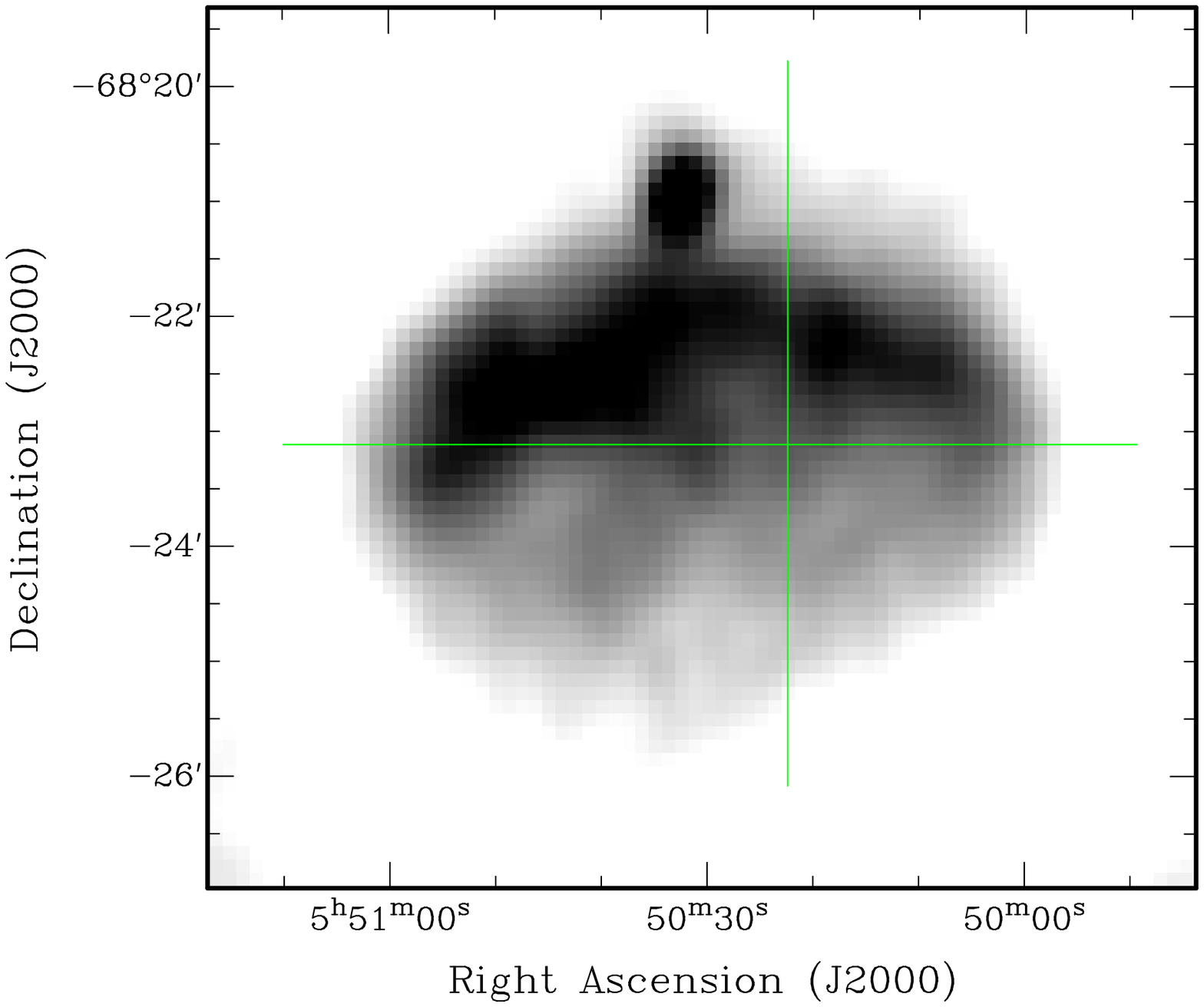}
 \includegraphics[angle=-90, width=110pt]{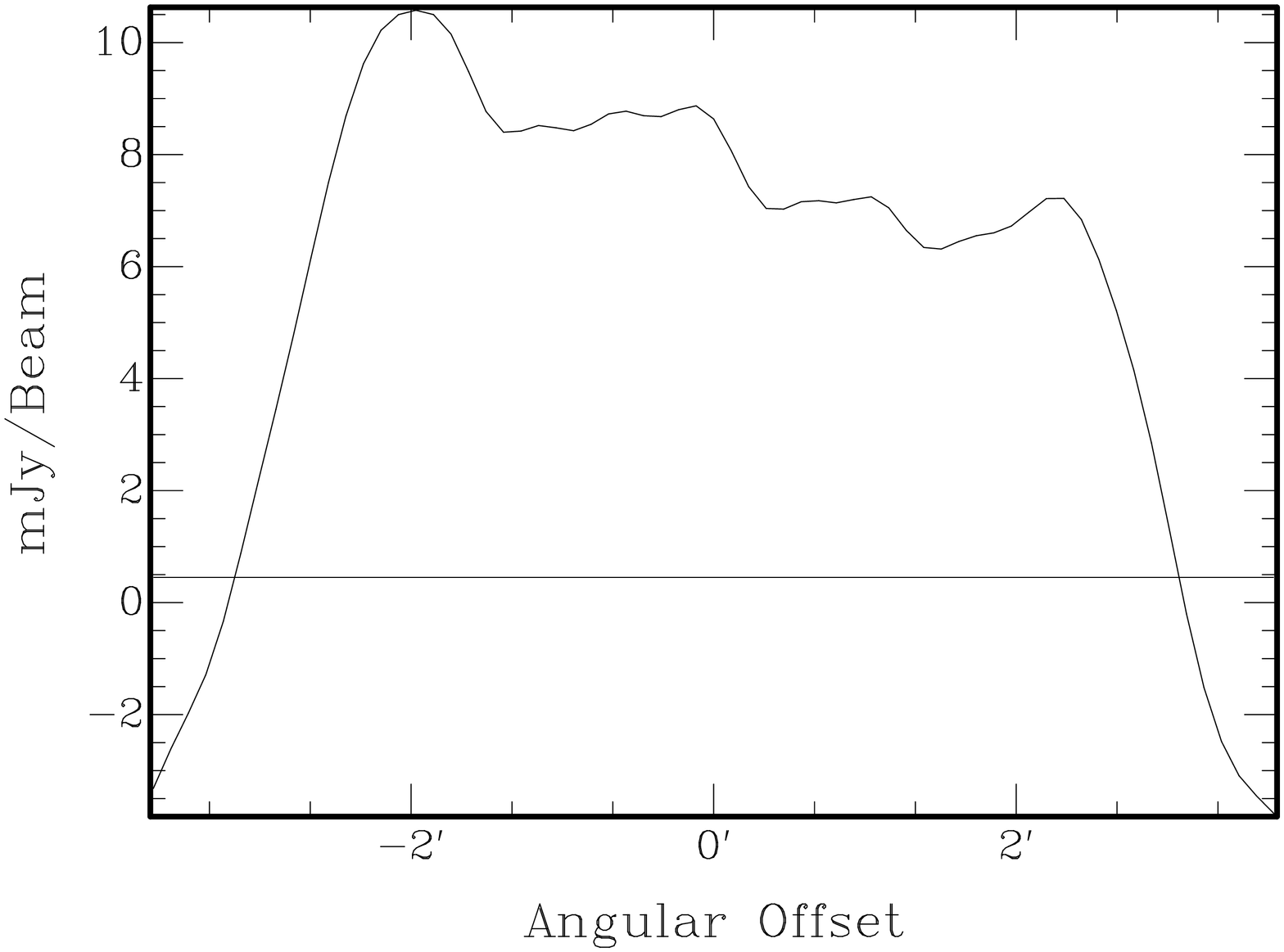}
 \includegraphics[angle=-90, width=110pt]{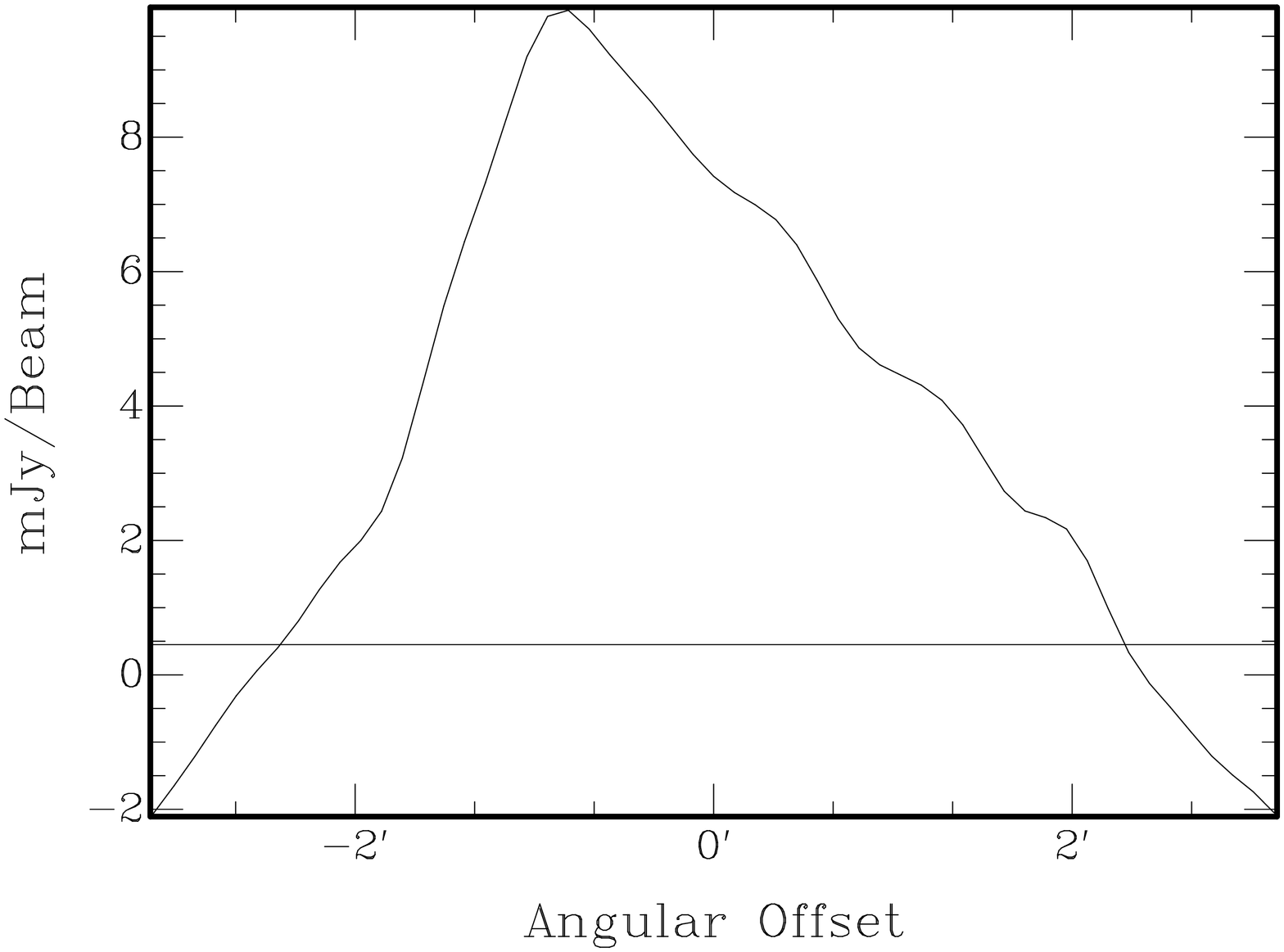}
  \caption{The left image shows the major and minor axis, with the major axis starting at the E end and the minor axis starting at the N end. The center image shows the I-Profile of the major axis with the 3$\sigma$ line shown. The right image shows the I-Profile of the minor axis with the 3$\sigma$ line shown.}
  \label{extent}
\end{figure}

SNR J0550-6823 exhibits a one sided shell brightened morphology a dissipating  in the southern region (Fig.~\ref{3cm6cont}). We note what is likely an unrelated background point source in its northern region. The remnant is centred at RA(J2000)=5$^h$50$^m$30.7$^s$, DEC(J2000)=--68\arcdeg23\arcmin37.0\arcsec\  with a diameter at 6~cm measuring $373\arcsec \times 282\arcsec \pm 4\arcsec\ $ ($90 \times 68 \pm 1$ pc). We estimate the extent at the 3$\sigma$ noise level (0.15~mJy) along the major (E--W) and minor (N--S) axis (PA=90\arcdeg) as presented in Fig.~\ref{extent}. However, we notice that at optical wavelengths SNR J0550-6823 extends further south and appears to have a near circular shape with the minor axis of $\sim$75-80~pc. Also, it appears that this SNR is more prominent in the [O\textsc{iii}] image and therefore is an excellent candidate for an Oxygen dominant type of SNR such as N\,132D or 1E0102-72. New upcoming observations similar to \citet{2011Ap&SS.331..521V} will confirm the true nature of this object.

\begin{figure}[h]
  \includegraphics[angle=-90, width=\columnwidth]{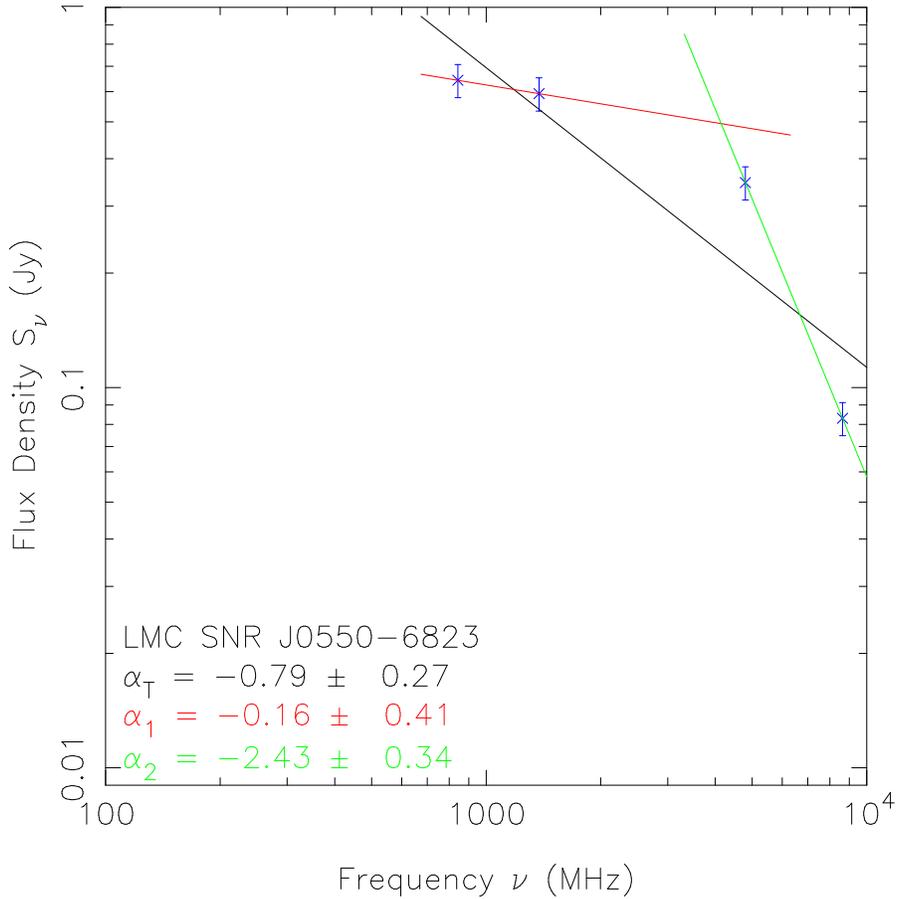}
  \caption{Radio-continuum spectrum of SNR J0550-6823. The 73~cm (408~MHz) value was disregarded in this estimate due to confusion from the strong point source in the northern region of the SNR.}
  \label{specindex}
\end{figure}

Using all values of integrated flux densities estimates (except for 73~cm; Table~1), a spectral index ($S\propto\nu^{\alpha}$) distribution is plotted in Fig.~\ref{specindex}. The overall radio-continuum spectra (Fig.~\ref{specindex}; black line) from SNR~0550-6823 was estimated to be $\alpha = -0.79 \pm 0.27$, while the typical SNR spectral index is $\alpha = -0.5 \pm 0.2$ \citep{1998A&AS..130..421F}. This somewhat steeper spectral index would indicate a younger age despite its  (large) size of $90 \times 68 \pm 1$~pc, suggesting it as an older (more evolved) SNR. We also note that this may indicate that a simple model does not accurately describe the data, and that a higher order model is needed. This is not unusual, given that several other Magellanic Clouds SNR's exhibit this ``curved'' spectra \citep{2008SerAJ.177...61C,2010SerAJ.181...43B}. Noting the breakdown of the power law fit at shorter wavelengths, we decomposed the spectral index estimate into two components, one ($\alpha_{1}$) between 36 and 20~cm, and the other ($\alpha_{2}$) between 6 and 3~cm. The first component (Fig.~\ref{specindex}; red line), $\alpha_{1} = -0.16 \pm 0.41$ is a reasonable fit and typical for an evolved SNR, whereas the second (Fig.~\ref{specindex}; green line), \mbox{$\alpha_{2} = -2.43 \pm 0.34$,} is a poor fit, and indicates that non-thermal emission can be described by different populations of electrons with different energy indices. Although the low flux at 3~cm (and to a lesser extent at 6~cm) could cause the large deviations, an underestimate of up to $\sim$50\% would still lead to a ``curved'' spectrum.  

SNR J0550-6823 is located on the eastern side of the LMC, far away from the main body of this dwarf galaxy. We also point out the dissipating shell in the southern region of the remnant. Therefore, it is reasonable to assume that this SNR is expanding in a very low density environment.

We estimate the spectral index of the point source (ATCA\,J0550-6820) in the northern region of the SNR to be $\alpha = -1.2 \pm 0.2$ (Table~1). This significantly steeper spectrum adds further evidence that the point source is unrelated to  SNR 0550-6823 and is most likely a background AGN. Hence, this background source may ``contaminate'' correct spectral index estimates of SNR~J0550-6823, especially in low-resolution studies such as \citet{1998A&AS..130..421F} ($-0.37 \pm 0.06$ previous vs. $-0.79 \pm 0.27$ this paper). For this reason we don't include the 73-cm flux density measurements in the new spectral index estimate, as the beam size (resolution) is over 2.5\arcmin. 

We also considered this point source to be a run-away pulsar related to SNR~0550-6823. We ruled out this scenario due to a lack of prominent trails (or pulsar wind nebulae) as found in LMC SNR\,N206 \citep{2002AJ....124.2135K} or SMC SNR~IKT\,16 \citep{2011A&A...530.A132}.

\begin{figure}[h]
  \includegraphics[angle=-90, width=\columnwidth]{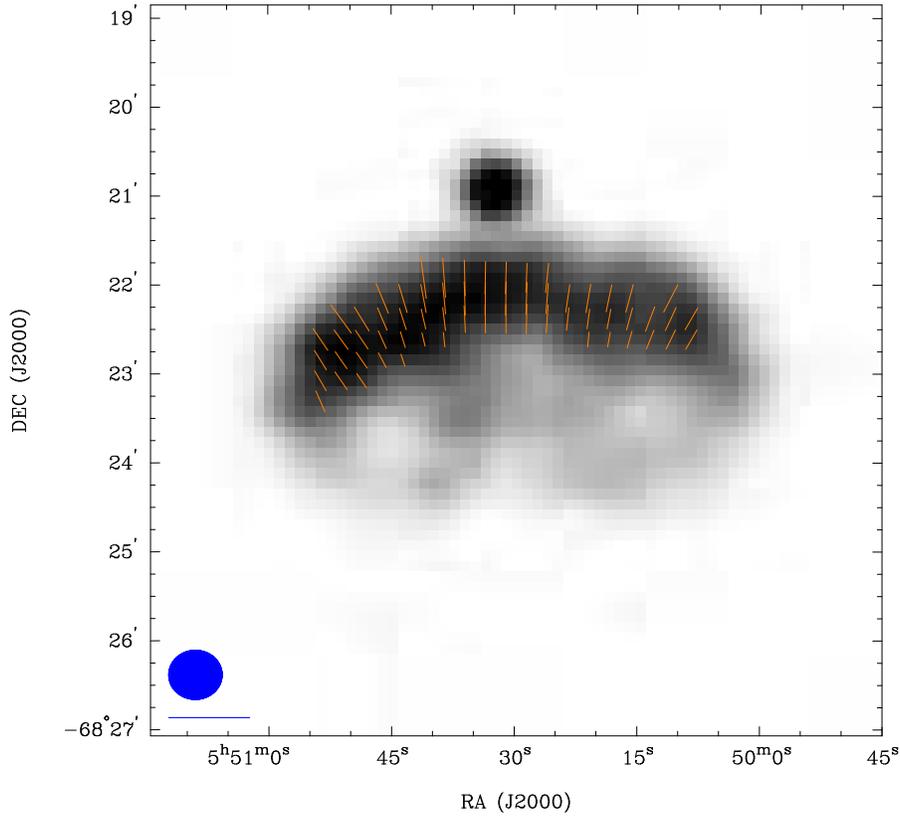}
  \caption{ATCA ``snap-shot'' observations (C634 project only) of SNR J0550-6823 at 6~cm (4.8~GHz) overlaid with polarisation vectors. The ellipse in the lower left corner represents the synthesised beamwidth of 36\,\arcsec$\times$33\arcsec, and the line below the ellipse is a polarisation vector of 100\%. }
  \label{polar}
\end{figure}

Linear polarisation images were created for each frequency using \textit{Q} and \textit{U} parameters (Fig.~\ref{polar}). Relatively strong linear polarisation is evident in the 6~cm image and is greater than many LMC SNRs \citep{2007MNRAS.378.1237B,2008SerAJ.176...59C,2008SerAJ.177...61C,2009SerAJ.179...55C,2010A&A...518A..35C} but somewhat weaker than  LMC SNR~J0527-6549 (DEM\,L204) \citep{2010SerAJ.181...43B}. 

The mean fractional polarisation at 6~cm was calculated using flux density and polarisation:
\begin{equation}
P=\frac{\sqrt{S_{Q}^{2}+S_{U}^{2}}}{S_{I}}\cdot 100\%
\end{equation}
\noindent where $S_{Q}, S_{U}$ and $S_{I}$ are integrated intensities for \textit{Q}, \textit{U} and \textit{I} Stokes parameters. Our estimated peak value at 6~cm is $50\% \pm10\%$ (Fig~\ref{polar}) while there is no reliable detection at 3~cm. Without reliable polarisation measurements at a second frequency, we could not determine the Faraday rotation and thus cannot deduce the magnetic field strength. We also note that the point source in the northern region is not polarised.  This is also consistent with it being an unrelated background source.

\section{Conclusion}

We carried out a radio-continuum study of SNR J0550-6823. From this analysis, we found that the SNR followed a one sided shell brightened morphology with a diameter of $373\arcsec \times 282\arcsec \pm 4\arcsec$ ($90 \times 68 \pm 1$ pc). it has a relatively flat spectral index ($\alpha  = -0.79 \pm 0.27$) and a strong 6~cm polarisation of $\sim 50\% \pm 10\%$. We also note correlations between the optical ([O\textsc{iii}]) and radio observation of this object, with the optical observations accounting for the seemingly ``missing'' southern emission seen in the images at radio wavelengths. These new observations will further improve our knowledge of this SNR as well as SNRs in general.\\
\\
We used the {\sc karma} software package developed by the ATNF. The Australia Telescope Compact Array is part of the Australia Telescope which is funded by the Commonwealth of Australia for operation as a National Facility managed by CSIRO. The Magellanic Clouds Emission Line Survey (MCELS) data are provided by R.C. Smith, P.F. Winkler, and S.D. Points. The MCELS project has been supported in part by NSF grants AST-9540747 and AST-0307613, and through the generous support of the Dean B. McLaughlin Fund at the University of Michigan, a bequest from the family of Dr. Dean B. McLaughlin in memory of his lasting impact on Astronomy. The National Optical Astronomy Observatory is operated by the Association of Universities for Research in Astronomy Inc. (AURA), under a cooperative agreement with the National Science Foundation. 

\clearpage
\vskip10mm

\begin{table}
\caption{Integrated Flux Density of SNR 0550-6823 and Point Source  ATCA\,J0550-6820}
\centerline{
\begin{tabular}{ccccccl}
\hline
$\nu$ & $\lambda$ & R.M.S  & Beam Size  & S$_\mathrm{SNR}$ & S$_\mathrm{PS}$ & Reference \\
(MHz) & (cm)      & (mJy)  & (arcsec)   & (mJy)            & (mJy)           &  \\
\hline
\p0408 & 73    & --     & 156$\times$156   & 980   & ---    & \citet{1976AuJPA..40....1C}\\
\p0843 & 36    & 1.5\p0 & 43.0$\times$43.0 & 643   &   132  & This Work\\
1377   & 20    & 1.5\p0 & 45.0$\times$45.0 & 593   & \p086  & This Work\\
4800   & \p06  & 0.15   & 40.2$\times$35.3 & 346   & \p024  & This Work\\
8640   & \p03  & 0.17   & 40.2$\times$35.3 & \p083 & \p0\p07& This Work\\
\hline
\end{tabular}}
\end{table}
%\vskip25mm

\clearpage

%It is also possible to automate the generation of the reference list
%itself using BibTeX, but support for this in the macros is still
%experimental. For further details, see \texttt{authorguide.pdf}.

%

\end{document}